\documentclass[published]{JHEP3}
\accepted{August 20, 2002}
\received{June 18, 2002}
\revised{August 16, 2002}
\JHEP{08(2002)033}

\usepackage{epsfig}

\skip\footins = 1\bigskipamount plus 2pt minus 6pt
\newcommand\SU{\mathop{\rm SU}}
\newcommand\U{\mathop{\rm {}U}}

\def\L{\mathcal{L}}
\def\d{\partial}
\def\sin{{\rm sin}}
\def\cos{{\rm cos}}
\def\tan{{\rm tan}}
\def\be{\begin{equation}} 
\def\ee{\end{equation}}
\def\bea{\begin{eqnarray}} 
\def\eea{\end{eqnarray}} 
\def\half{\frac{1}{2}}
\def\quarter{\frac{1}{4}}
\def\tw{{\theta_{{{\rm W}}}}}

\title{The evolution and persistence of dumbbells} 

\author{Jon Urrestilla \\
Department of Theoretical Physics, University of the Basque Country
Bilbao, Spain, and\\
Astronomy Centre, University of Sussex
Falmer, Brighton BN1 9QJ, United Kingdom\\
E-mail: \email{wtbururj@lg.ehu.es} }

\author{Ana Ach\'{u}carro\\
Lorentz Institute of Theoretical Physics, University of Leiden\\
2333 RA Leiden, The Netherlands, and\\
Department of Theoretical Physics, University of the Basque Country
Bilbao, Spain, and\\
Institute for Theoretical Physics, University of Groningen
The Netherlands\\
E-mail: \email{a.achucarro@phys.rug.nl} }

\author{Julian Borrill\\
National Energy Research Scientific Computing Center
Lawrence Berkeley National Laboratory, University of California
Berkeley, CA 94720, USA, and\\
Center for Particle Astrophysics, University of California
Berkeley, CA 94720, USA\\
E-mail: \email{borrill@nersc.gov}}

\author{Andrew R.~Liddle\\
Astronomy Centre, University of Sussex
Falmer, Brighton BN1 9QJ, United Kingdom\\
E-mail: \email{a.liddle@sussex.ac.uk} }

\abstract{We use large-scale numerical simulations to study the
formation and evolution of non-topological defects in a generalized
electroweak phase transition described by the Glashow-Salam-Weinberg
model without fermions. Such defects include dumbbells, comprising a
pair of monopoles joined by a segment of electroweak string. These
exhibit complex dynamics, with some shrinking under the string tension
and others growing due to the monopole-antimonopole attractions
between near neighbours. We estimate the range of parameters where the
network of dumbbells persists, and show that this region is narrower
than the region within which infinite straight electroweak strings are
perturbatively stable.}

\begin{document}

\section{Introduction}

The formation of topological defects in any phase transition depends
on the existence of a non-trivial, low-order homotopy group of the
broken-symmetry vacuum manifold~\cite{K76,VS}. Even in the absence of
this, dynamically stable non-topological defects are sometimes still
possible, but are usually assumed to be too weakly stable to lead to a
lasting network. A particularly interesting case is non-topological
strings where at least one example is known (semilocal
strings~\cite{VA91,H92}) in which a space-spanning network of strings
can develop from the growth and joining of short string segments
formed at the phase transition~\cite{ABL99}.  Semilocal strings are a
special case of electroweak strings~\cite{N77,V92,report}.

The stability of electroweak strings has been analyzed in some detail
in the Glashow-Salam-Weinberg (GSW) model.  In the absence of fermions
the theory is determined, up to scalings, by only two parameters: the
weak mixing angle $\tw$, and $\beta$ which is the square of the ratio
of the Higgs and $Z$-boson masses (the observed value for the actual
electroweak model is $\sin^2\tw \simeq 0.23$, and the precise value of
$\beta$ has not yet been determined but is almost certainly greater
than one).

In the case of interest here, the topology of the vacuum manifold is
the three-sphere $S^3$, which does not support persistent topological
defects in 3+1 dimensional spacetime, but there are grounds for
expecting non-topological defects to form. In the limit $\tw = \pi/2$
we recover the so-called semilocal model, whose $S^3$ vacuum manifold
is known to support stable non-topological strings from both
analytical~\cite{VA91,H92} and numerical~\cite{ABL99} work. In
particular, a straight infinite semilocal string can be seen as a
Nielsen-Olesen (NO) $\U(1)$ vortex~\cite{NO}, embedded in a higher
group $\SU(2)_{{\rm global}}\times \U(1)_{{\rm local}}$. Note,
however, that in the semilocal string case, we can define a quantity
(the winding number) which, although not a topological invariant in
the usual sense, is topologically conserved.

The embedding of NO vortices into the full electroweak symmetry
$\SU(2)_L\times \U(1)_{{\rm{\scriptscriptstyle Y}}}$ leads to
$Z$-strings and $W$-strings (see ref.~\cite{report}). Although
$W$-strings are expected to be unstable, analysis of infinite
axially-symmetric $Z$-strings has shown the existence of a parameter
regime where they are perturbatively stable~\cite{V92,JPV93}. But
these $Z$-strings are genuinely non-topological, there is no quantity
which is topologically conserved.

In this paper we will consider the evolution and persistence of these
genuine non-topological defects in a generalized GSW model, spanning
all values of $\beta$ and $\tw$, using numerical methods.  Isolated,
infinite, axially-symmetric $Z$-string configurations will not be
formed in a realistic system, but configurations consisting of
monopole-antimonopole pairs joined by $Z$-string segments, named
\emph{dumbbells} by Nambu~\cite{N77}, are certainly possible.
Isolated dumbbells are expected to collapse under the string tension,
at least in the absence of rotation, jittering or magnetic
fields~\cite{N77,V02,GM95}.  Of particular interest is the question of
whether there is a region of parameter space in which the density of
dumbbells is sufficiently high so that they are able to generate a
persistent network of strings by building up longer segments from the
merging of shorter ones due to interaction of neighbouring monopoles,
as has been found in the semilocal case~\cite{ABL99}.  This region of
parameter space will lie far from the measured values for $\tw$ and
$\beta$, but the mere existence of this region is striking.  It shows
that in models close to real physical ones, completely non-topological
defect networks can form and persist, so in extensions of the GSW
model (to higher symmetry groups or extra fields), or models with a
background magnetic field~\cite{GM95}, or even models in which
topological (semilocal) defects exist in some limit, non-topological
defects cannot be ruled out immediately.

Moreover, the study of the dynamics of dumbbells can be of interest in
computing other early universe features, such as the primordial
magnetic field helicity~\cite{V01,VF94}.

\section{The model}

The bosonic sector of the GSW electroweak model describes an $\SU(2)_L
\times \U(1)_{{\rm{\scriptscriptstyle Y}}}$ invariant theory with a
scalar field $\Phi$ in the fundamental representation of $\SU(2)_L$,
with lagrangian
\be 
\L = \left|D_\mu\Phi\right|^2-\quarter
W^a_{\mu\nu}W^{a\mu\nu}-\quarter
Y_{\mu\nu}Y^{\mu\nu}-\lambda\left(\Phi\Phi^\dagger-\frac{\eta^2}{2}\right)^2
.
\ee 
The covariant derivative is given by 
\be
D_\mu\equiv\d_\mu-\frac{ig_{{\rm {\scriptscriptstyle 
W}}}}{2}\tau^aW^a_\mu-\frac{i g_{{\rm {\scriptscriptstyle Y}}}}{2}
Y_\mu  \, , \qquad a=1,2,3 \,,
\ee 
where $\Phi$ is a complex doublet, $\tau^a$ are the Pauli matrices,
$W_\mu^a$ is a $\SU(2)$ gauge field and $Y_\mu$ is a $\U(1)$ gauge
field. The field strengths associated with these gauge fields are
\bea
W_{\mu\nu}^a&=&\d_\mu W^a_\nu-\d_\nu W^a_\mu+g_{{\rm{\scriptscriptstyle W}}}
\epsilon^{abc}W_\mu^bW_\nu^c \,;
\nonumber\\
Y_{\mu\nu}&=&\d_\mu Y_\nu-\d_\nu Y_\mu \,,
\eea 
respectively, and there is no distinction between upper and lower
group indices ($\epsilon^{123}=1$).

When the scalar field acquires a non-zero vacuum expectation value the
symmetry breaks from $\SU(2)_L\times \U(1)_{{\rm {\scriptscriptstyle
Y}}}$ to $\U(1)_{{\rm e.m.}}$, leaving a massive scalar field
($m_H=\sqrt{2\lambda}\eta$), a massless neutral photon ($A_\mu$), a
massive neutral $Z$-boson ($Z_\mu$, $m_{{{\rm z}}}=g_{{{\rm z}}}\eta
/2 \equiv l_v^{-1}$, where $g_{{{\rm z}}}=\sqrt{g_{{\rm
{\scriptscriptstyle Y}}}^2+g_{{\rm {\scriptscriptstyle W}}}^2}$), and
two massive charged $W$-bosons ($W_\mu^\pm$, $m_W=g_{{\rm
{\scriptscriptstyle W}}}\eta /2$).

We make the following rescaling
\be
\Phi\to\frac{\eta}{\sqrt{2}}\Phi\,,\qquad
x_\mu\to\frac{\sqrt{2}}{g_{{{\rm z}}} \eta} x_\mu \,, \qquad 
g_{\rm {\scriptscriptstyle Y}}Y_\mu\to\frac{g_{{{\rm
z}}}\eta}{\sqrt{2}}Y_\mu \,, \qquad 
g_{\rm{\scriptscriptstyle W}} W^a_\mu\to\frac{g_{{{\rm
z}}}\eta}{\sqrt{2}}W^a_\mu \,,
\label{resc}
\ee
to choose $l_v$ as the unit of length, $\eta$ as the unit of energy,
and the $Z$-charge of the scalar field ($g_{{{\rm z}}}$) as the unit
of charge (up to numerical factors). This brings the classical field
equations to the form
\bea
D^\mu D_\mu\Phi+\frac{2\lambda}{g_{{{\rm 
z}}}^2}\left(\Phi\Phi^\dagger-1\right)\Phi&=&0\,; 
\nonumber\\
\d_\nu W^{\mu\nu a}+\epsilon^{abc}W_\nu^b W^{\mu\nu 
c}&=&\frac{i}{2}\cos^2\,\tw\left[\Phi^\dagger\tau^aD^\mu 
\Phi-\left(D^\mu\Phi\right)^\dagger\tau^a\Phi\right]; 
\nonumber\\
\d_\nu Y^{\mu\nu}&=&\frac{i}{2}\sin^2\,\tw\left[\Phi^\dagger 
D^\mu\Phi-\left(D^\mu\Phi\right)^\dagger\Phi\right],
\label{eom}
\eea
where the weak mixing angle is given by $\tan\,\tw\equiv g_{{\rm
{\scriptscriptstyle Y}}}/g_{{\rm {\scriptscriptstyle W}}}$, and now
\bea
W_{\mu\nu}^a&\equiv&\d_\mu W_{\nu}^a-\d_\nu W_{\mu}^a+
\epsilon^{a b c}W_{\mu}^b W_{\nu}^c\,; 
\nonumber\\
D_\mu&\equiv&\d_\mu-\frac{i}{2}\tau^a 
W_\mu^a-\frac{i}{2}Y_\mu\,.
\eea
Typically the $Z$- and $A$-fields are expressed in the unitary gauge
$\Phi^T=(0,1)$, but this choice is not well suited to work with
defects. Instead, when there are points in space-time with $|\Phi|\ne
1$, it is customary to use a more general definition of these fields
which depends on the Higgs field configuration at each
point~\cite{N77}, namely,
\bea
Z_\mu & \equiv & \cos\,\tw\,n^a(x)\,W^a_\mu-\sin\,\tw\,Y_\mu \,; 
\nonumber\\
A_\mu & \equiv & \sin\,\tw\,n^a(x)\,W^a_\mu+\cos\,\tw
Y_\mu   \,, 
\label{fie1}
\eea 
where 
\be
n^a(x)\equiv-\frac{\Phi^\dagger(x)\tau^a\Phi(x)}{\Phi^\dagger(x)\Phi(x)} \,,
\label{na}
\ee 
is a unit vector by virtue of the Fierz identity $\sum_a \left(
\Phi^\dagger \tau^a \Phi \right)^2 = \left( \Phi^\dagger \Phi
\right)^2$. There are also several possible definitions for the field
strengths (see, e.g.~ref.~\cite{H94} for a discussion of this point);
in our simulations we will use
\bea
Z_{\mu\nu}&=&\cos\tw\,n^a(x)\,W_{\mu\nu}^a-\sin\tw\,Y_{\mu\nu} \,; 
\nonumber\\
A_{\mu\nu}&=&\sin\tw\,n^a(x)\,W_{\mu\nu}^a+\cos\tw\,Y_{\mu\nu}\,.
\label{fieldstrength}
\eea 
Note that eqs.~(\ref{fie1}) to~(\ref{fieldstrength}) reduce to the
usual definitions away from the defect cores.  We work in flat space
and in the temporal gauge ($W_0^a=Y_0=0$ for $a=1,2,3$) so
$D_0\Phi=\d_0 \Phi$. With this gauge choice, Gauss's Law becomes 
\bea
-\d_j(\d_0 Y_j) & = & \frac{i}{2}\sin^2\tw \left[\Phi^\dagger\d_0
\Phi-(\d_0\Phi)^\dagger\Phi\right]; 
\nonumber\\ 
-\d_j(\d_0 W_j^a)
-\epsilon^{abc}W_j^b\d_0W_j^c & = & \frac{i}{2} \cos^2\tw
\left[\Phi^\dagger\tau^a\d_0\Phi-(\d_0\Phi)^\dagger\tau^a\Phi\right], 
\eea 
(with $j=1,2,3$) which is then used to test the stability of the code.
 
As noted above, the $\sin^2\tw=1$ case reduces to the semilocal case,
where absolutely stable defects are known to exist. Setting one of the
Higgs fields to zero further reduces this to the abelian Higgs
model. We are able to use our knowledge of these systems, as well as
the behaviour of artificially constructed infinite axially-symmetric
$Z$-strings at various points in parameter space, to check the
validity of our simulation code.

\section{Numerical simulations}
\label{ns}

The equations of motion eq.~(\ref{eom}) are discretized using a
na\"{\i}ve staggered leapfrog method, where both the scalar and gauge
fields are associated with lattice points.  This procedure is the
easiest to relate to previous work on semilocal defects~\cite{ABL99},
and we expect to have a fairly good picture at the scales we are
interested in.  A comparison to a link-variable
discretization~\cite{KS} of eq.~(\ref{eom}) in the semilocal case
($\tw=\pi/2$) shows that the changes from using a lattice
implementation instead of this na\"{\i}ve one are well within the
other uncertainties; see the appendix for a full analysis.  The
simulations are performed on a periodic cubic lattice whose time step
is $0.2$ times the spatial step $\Delta t = 0.2 \Delta x$ ($c=1$),
with \emph{ad hoc} numerical viscosity terms added to each equation
($\gamma\,\dot\Phi$, $\gamma\,\dot Y$ and $\gamma\,\dot W^a$
respectively) to reduce the system's relaxation time.  The expansion
rate in an expanding universe would play this role, albeit as a
time-dependent factor $\gamma(t)$ typically scaling as $1/t$.  Several
different values of $\gamma$ were tested, and generated very similar
behaviour; throughout this work we use $\gamma=0.5$.

Two different strategies for setting the initial configurations were
considered:
\begin{itemize} 
\item[$(a)$] Set all initial field velocities to zero and throw
down random scalar field phases at every lattice point; average the
field at each point with its 6 nearest neighbours and normalize the
fields to the vacuum value ($|\Phi|^2=1$ as $\eta$ has been rescaled
out) iteratively (50 times), to get a smoother configuration; using
these scalar field values, choose the initial values of the gauge
fields to be
\bea
Y_\mu&=&0\nonumber\\
W_\mu^1&=&
2\left(\psi_1\nabla_j\psi_4-\psi_4\nabla_j\psi_1
+\psi_3\nabla_j\psi_2-\psi_2\nabla_j\psi_3\right)
\nonumber\\
W^2_\mu&=&
2\left(\psi_3\nabla_j\psi_1-\psi_1\nabla_j\psi_3
+\psi_4\nabla_j\psi_2-\psi_2\nabla_j\psi_4\right)
\nonumber\\
W^3_\mu&=& 2\left(\psi_1\nabla_j\psi_2-\psi_2\nabla_j\psi_1
+\psi_4\nabla_j\psi_3-\psi_3\nabla_j\psi_4\right)
\label{mini}
\eea 
\looseness=-1 where $\Phi^T=(\psi_1+i\,\psi_2,\psi_3+i\,\psi_4)$,
pseudo-minimizing the energy as in ref.~\cite{ABL98}. Note, however,
that in the present case, the system does have enough gauge fields to
cancel completelly the gradient energy. In fact, the
configuration~(\ref{mini}) sets to zero the gradient energy, the
potential energy and the field-strength energy corresponding to the
$\U(1)$ gauge field, leaving the $\SU(2)$ field-strength energy as the
only non-zero term.

\pagebreak[3]

\item[$(b)$] Set all fields to zero, and also gauge field velocities to zero
($\Phi=W^a_\mu=\dot W^a_\mu=Y_\mu=\dot Y_\mu=0$); provide some initial
(smoothed) random velocities to the scalar field ($\dot\Phi\ne 0$)
following the general procedure as above. This initial configuration
would be closer to that appropriate to a bubble nucleation scheme.
\end{itemize}

The overall results in our simulations did not show qualitatively
different behaviour between the two cases.  Indeed, in the first case
the gauge field energy pseudo-minimization was ineffective enough that
the scalar field typically began by climbing up the potential,
restoring the symmetric phase, and rolled down again later.  Previous
experience with semilocal strings also shows that the results are
fairly insensitive to the specific way initial conditions are
implemented.  For all these reasons, we chose the second initial
configuration ($b$) for our investigations.

\looseness=1 Interpreting simulations of electroweak string networks
is more complicated than in the $\U(1)$ cosmic string case. As in the
semilocal case, electroweak strings are non-topological, and in this
case there is no well-defined winding number, which makes
identification of strings a more difficult task. We follow the
strategy proposed in ref.~\cite{ABL98} in order to study string
formation: the system is evolved forward in time and we compute a set
of gauge-invariant quantities at each time step, namely the $Z$- and
the $A$-field strengths given by eq.~(\ref{fieldstrength}) and the
modulus of the scalar field ($|\Phi|$). String formation is then
observed by visualizing isosurfaces in the $Z$-field strength
($\sqrt{\half Z_{ij}Z^{ij}}$) and in the scalar field modulus.

After the rescaling eq.~(\ref{resc}), it becomes clear that the only
free parameters in our model are $\beta$ ($\beta \equiv m_H^2/m_{{{\rm
z}}}^2=8\lambda/g_{{\rm z}}^2$) and $\tw$.  It is well known that with
$\tw=\pi/2$ and $\beta<1$ we have semilocal strings that are
stable~\cite{H92}, and numerical simulations show segment formation
and linkage~\cite{ABL99}. Moreover, for $\sin^2\tw\lesssim 1$,
$\beta<1$, there is a regime, albeit a rather narrow one, where
infinitely long strings are perturbatively stable~\cite{V92,JPV93}.
Bearing these results in mind, the parameter space investigated in our
simulations was $0.9 \leq \sin^2\tw \leq 1$ and $0.05 \leq \beta\leq
1.5$.

Simulation and visualization were performed on $64^3$, $128^3$ and
$256^3$ lattices using the Cray T3E at NERSC, and high-performance
computing facilities at University of Sussex and University of
Groningen. Our quoted results all come from $256^3$ simulations.

\section{Results}
\label{results}

To test our code, we began by reproducing known results for cosmic and
semilocal string networks, and for infinitely long, axially-symmetric
$Z$-strings (made possible by our periodic boundary conditions as
long as strings are simulated in pairs to keep the net flux equal to
zero) in both the stable and unstable regimes, in particular verifying
that the $Z$-strings disappeared in the unstable regime.

Having checked the code, we then ran it for values of the two free
parameters ($\theta_W, \beta$) throughout the regime of interest,
using the same initial conditions for each parameter pair. We chose to
carry out one large simulation at each point in parameter space rather
than many smaller ones, which gives improved dynamical range. As
expected, after an initia\pagebreak[3]l transient, the typical configuration
observed is a dumbbell --- a segment of electroweak string joining a
monopole/antimonopole pair. By continuity, for parameter values
sufficiently close to the stable semilocal case we expect the monopole
interactions to lead to the joining of strings to form longer
segments.

{\renewcommand{\belowcaptionskip}{-6pt}
\FIGURE[t]{
\epsfig{file=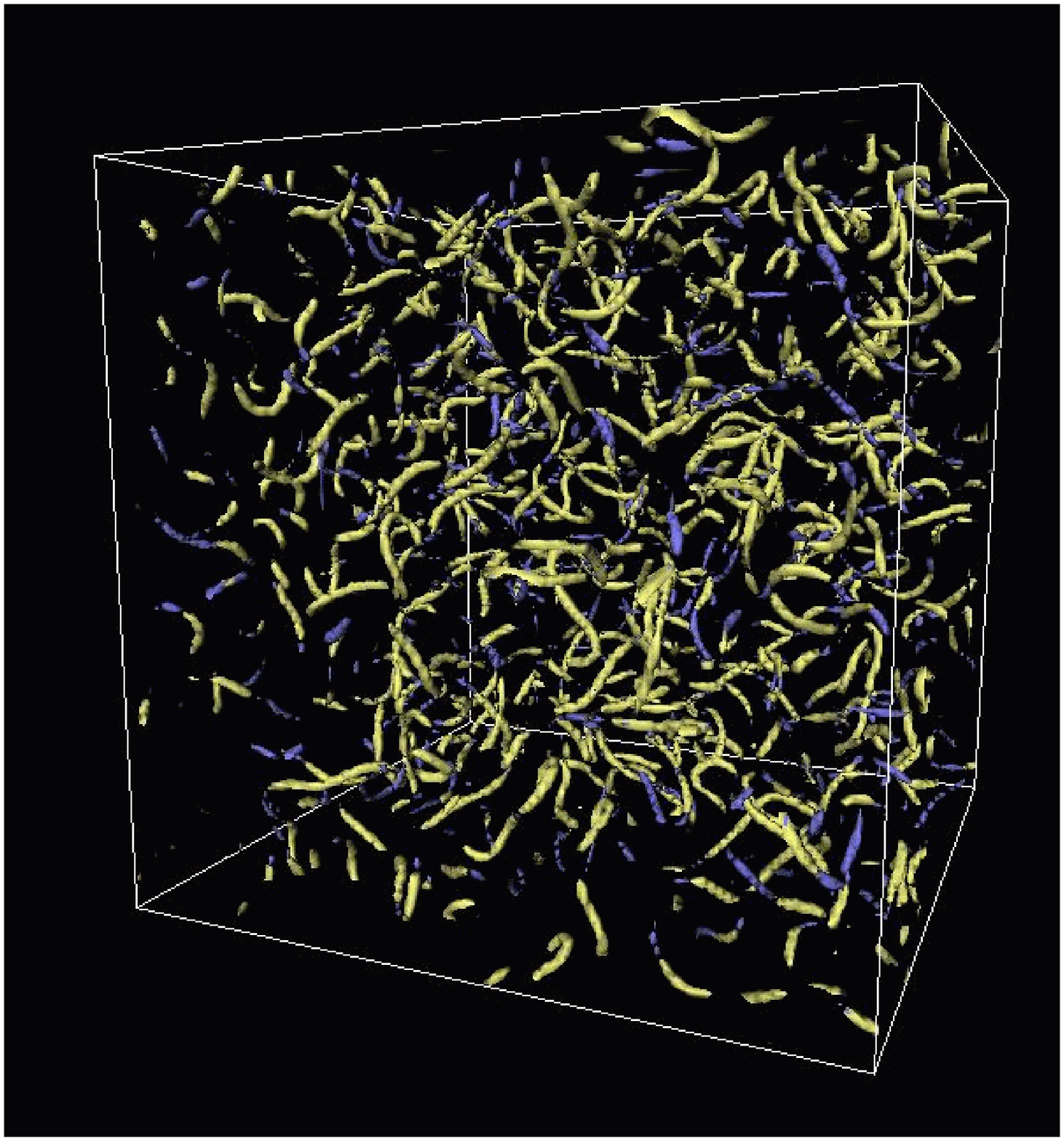, width=7cm}
\epsfig{file=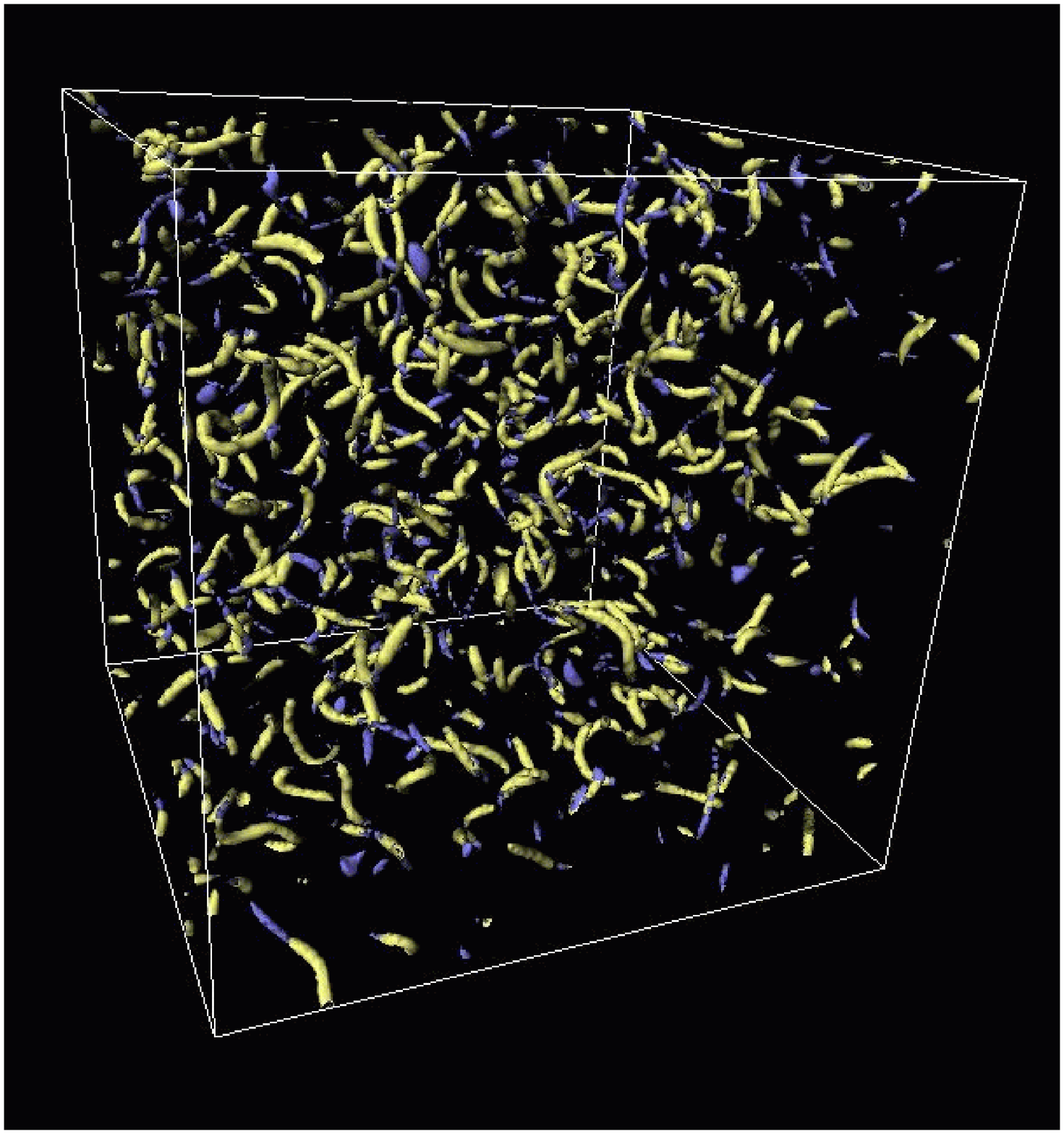, width=7cm}\\[2pt]
\epsfig{file=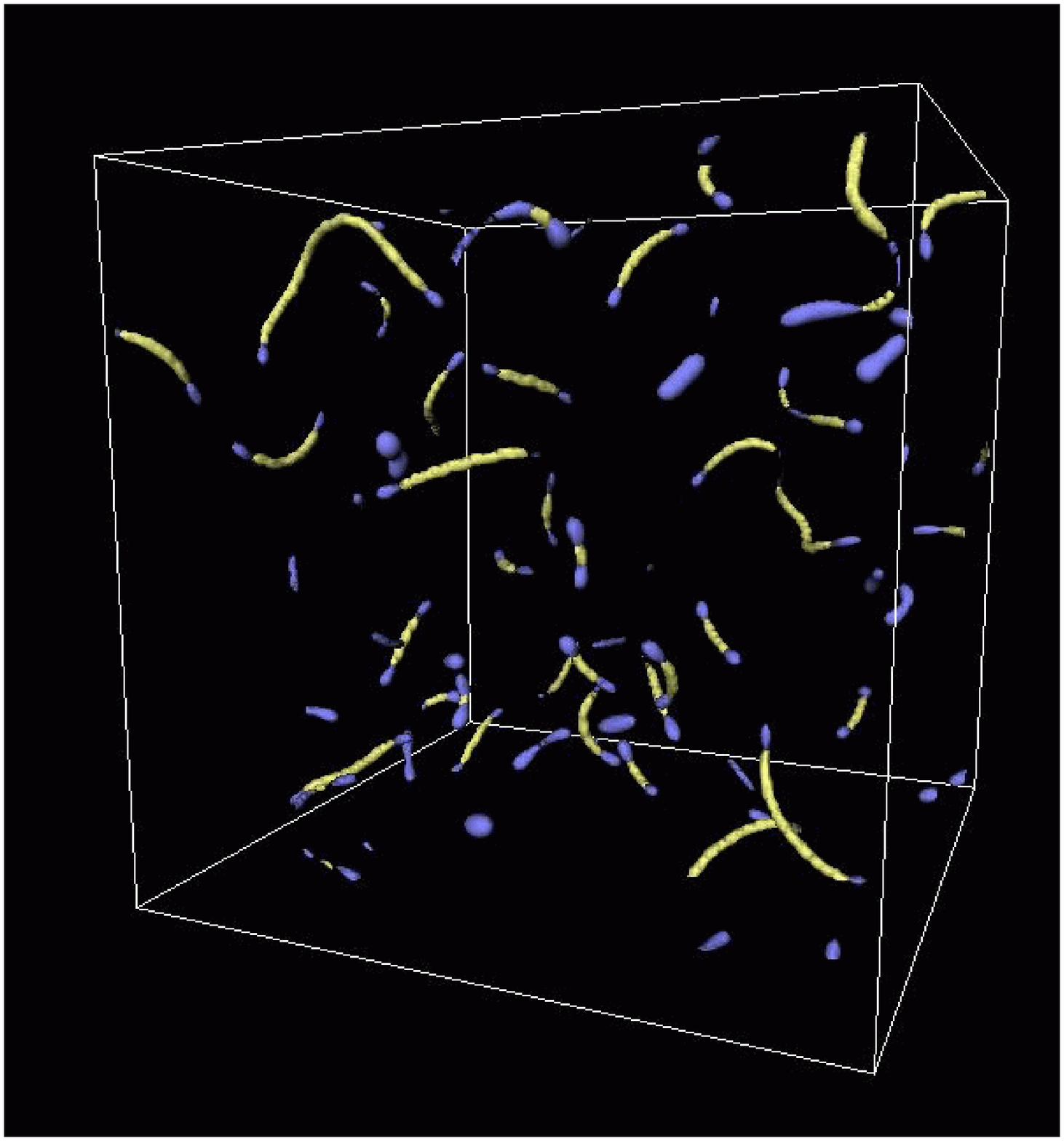, width=7cm} 
\epsfig{file=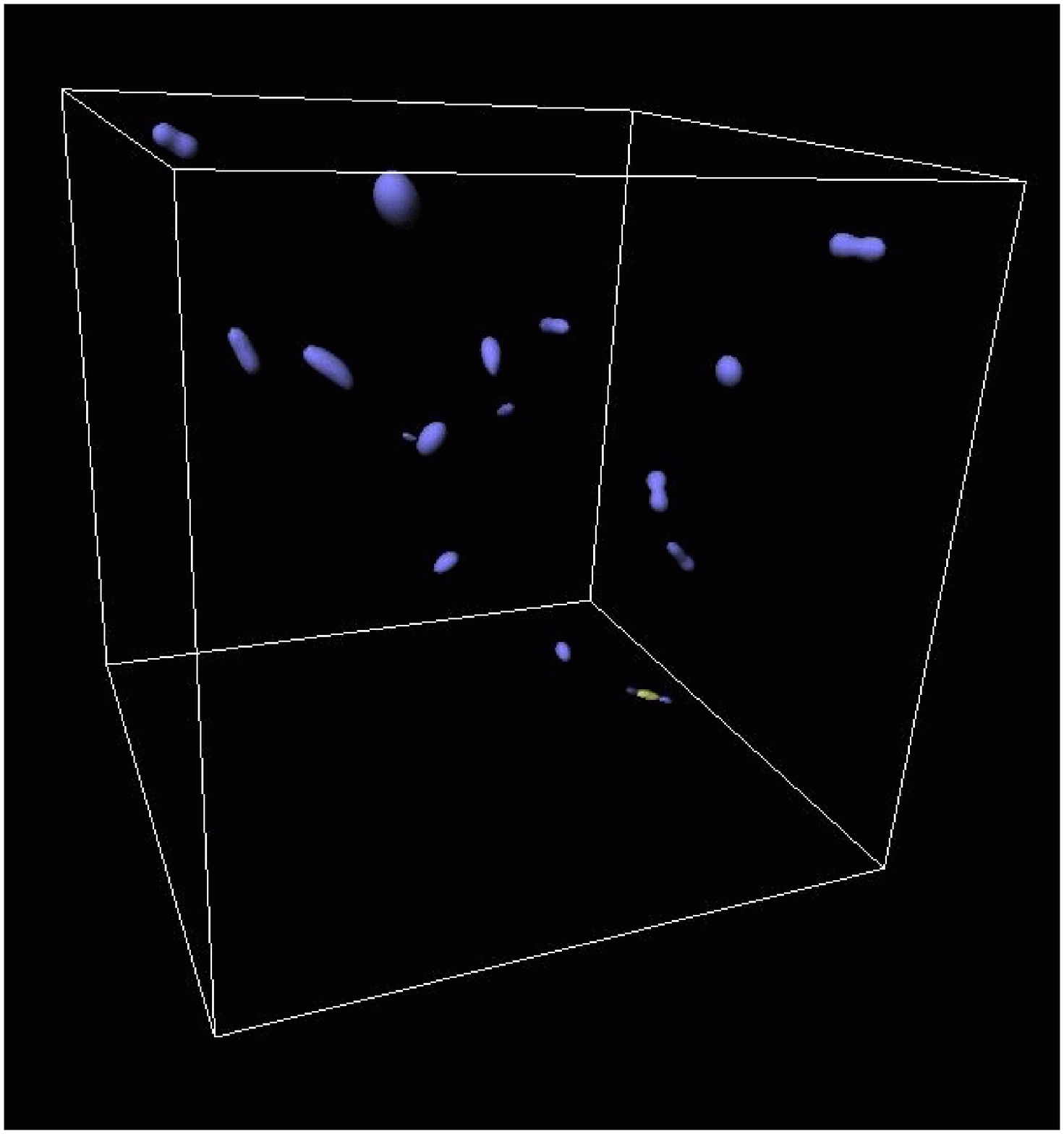, width=7cm}
\caption{\label{dumbini}\label{dumbfin}Isosurfaces of the
$Z$- and $A$-magnetic field strengths for two different simulations,
shown as the light and dark colour respectively. The left panels show
$\beta=0.1$, $\sin^2\tw=0.994$ (persistent regime) and the right ones
$\beta=0.5$, $\sin^2\tw=0.995$ (non-persistent regime). The top row
is at an early stage of the simulation $t=50$, while the lower is at
the end $t=200$. Note that in the first case there remain some long
strings at the end of the simulation and connection can still
occur. In the second all the defects are about to disappear.}}}

The simulations show that some short segments of $Z$-string do join as
expected. The joining rate is, however, lower than in the semilocal
case, and decreases both as $\sin^2\tw$ is decreased and as $\beta$ is
increased. Thus, some segments which would eventually join in the
semilocal case are seen instead to collapse in the electroweak case
due to string tension.  This is not surprising: in the semilocal case,
we have global monopoles at the string ends, which have divergent
scalar gradient energy and are more efficient at finding neighbouring
monopoles. In the electroweak case, the monopoles at the string ends
are proper magnetic monopoles, and the scalar gradients are cancelled
much more efficiently by gauge fields. As $\sin^2\tw \to 1$ the cores
become larger, and eventually overlap, making segments join. But as we
move away from the semilocal case, the cores become smaller and the
joining becomes less important.

Figure~\ref{dumbfin} shows timeslices of two typical simulations, with
the different colours corresponding to the $A$- and $Z$-field
strengths.\footnote{Further colour images and movies can be found
at\newline
\href{http://www.nersc.gov/~borrill/defects/electroweak.html} {\tt
http://www.nersc.gov/$\sim$borrill/defects/electroweak.html}}

The $Z$-field has a string-like form, whereas the $A$-field at the
string ends is a spherical shell, corresponding to spherical magnetic
monopoles. The $A$-field morphology can also be tube-like, denoting
interaction between monopoles, illustrating the complexity of the
overall dynamics. The interplay between string tension and
monopole-antimonopole attraction causes some strings to shrink until
they disappear, and others join to form longer strings.

To compare the number of defects in the semilocal and electroweak
cases, we plot the number of lattice sites with $Z$-magnetic field
strength ($\sqrt{\half Z_{ij}Z^{ij}}$) greater than 25\% of the
maximum field strength found in the core of a NO string with the same
value of $\beta$ (see figure~\ref{points}). The number of lattice
sites, and hence length of string, decreases more rapidly in the
electroweak case as either $\sin^2\tw$ decreases or $\beta$ grows.

{\renewcommand{\belowcaptionskip}{-10pt} 
\EPSFIGURE[t]{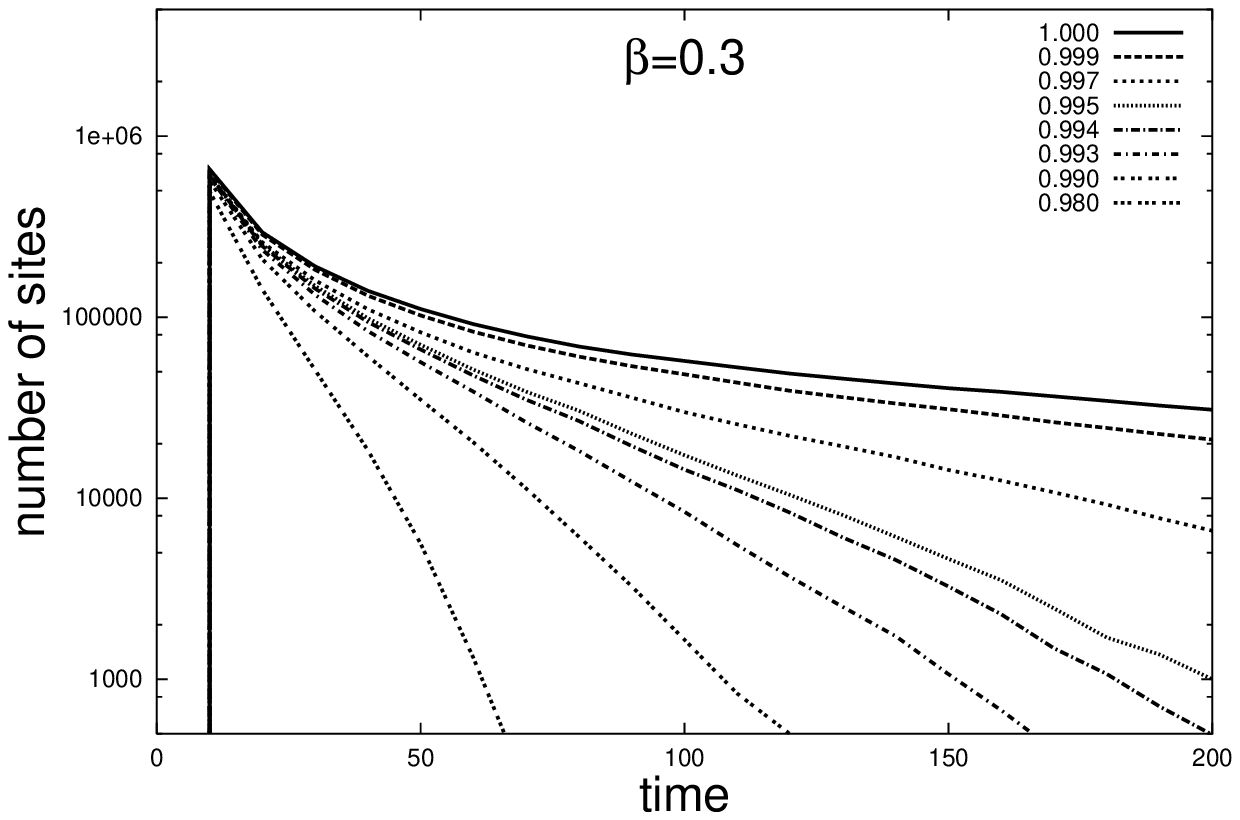,width=0.8\textwidth}{\label{points}The number of
lattice points with $Z$-field strength ($\sqrt{\half Z_{ij}Z^{ij}}$)
bigger than $25\%$ of the maximum (core) value in the NO case,
calculated in a series of $256^3$ simulations, for $\beta=0.3$.  The
different lines correspond to different choices of the weak mixing
angle $\tw$ (values shown are for $\sin^2\tw$). Note that in our
criterion for persistence (see text), for the chosen $\beta$, the
defects are considered to live long enough for $\sin^2\tw \gtrsim
0.995$. Different initial conditions for the simulations give similar
behaviour.}}

At early times the configurations in different simulations are very
similar. During the symmetry-breaking transient (the very first time
steps) no defects exist. As the scalar field takes on a non-zero value
a large number of very small string segments emerge, then the magnetic
monopoles become visible at the ends of the segments due to the
gathering of $A$-magnetic flux there.  The upper panels of
figure~\ref{dumbini} show the $Z$- and $A$-magnetic field strength of
two different simulations at time $t=50$. Similarities in these first
time steps, together with similarities using different initial
conditions (see above), show that the initial configurations (i.e.~the
way the symmetry breaking is implemented) are not as important as the
subsequent interaction between the scalar and gauge fields.

\looseness=-1 As the system evolves, we see that in only one of the
two simulations do the small segments grow and connect to their
neighbours, allowing the defect network to persist. In the lower
panels the configuration at time $t=200$ can be seen. In one case
there are long strings and connections are still happening, whereas in
the other almost all of the dumbbells have annihilated. In particular,
in the first case the final configuration contains strings which are
much longer at the end of the simulation than any present in the early
stages.

We wish to determine the parameter space for which the string network
persists. The stability of the corresponding infinite $Z$-strings is a
necessary condition for persistence, but not sufficient since the
physically realizable collection of finite initial string segments may
not link up. We therefore expect the persistence region of parameter
space to lie entirely within the stability region.

To estimate the persistence of $Z$-strings we need to define a
criterion to decide when defects are lasting long enough. There is
clearly some degree of arbitrariness in how this criterion is set, and
we have been guided by inspecting visually the evolution in different
parameter regimes. In order to construct the persistence region in
parameter space we simulated the system in $256^3$ boxes with periodic
boundary conditions, and considered persistence to have occurred if at
time $t=200$ there are more than 1000 lattice sites with a
$Z$-magnetic flux greater than $25\%$ of the maximum of the NO
simulation for the same $\beta$. For example, for $\beta = 0.3$, we
can see in figure~\ref{points} that persistence defined this way is
exhibited only for $\sin^2 \tw > 0.995$.  From a suite of simulations
using this criterion we obtain the `persistence limit' shown in
figure~\ref{persist}. As anticipated, the persistence region covers
only a subset of the stability region, with persistence only for
values of $\sin^2 \tw$ extremely close to one.  Another possible
criterion for persistence which can be easily automated is the
presence at late times of strings that are reasonably long compared to
their width.\footnote{\label{automate}To automate the calculation of
the length of the individual strings, we compute the volume of each
segment of string by counting the number of connected points whose
$Z$-magnetic field is at least 25\% of the maximum in a NO string at
the same value of $\beta$. Then, the volume is divided by the
cross-sectional area of the NO string.  Finally, to obtain the
length-to-width ratio, we divide it by the diameter of the
corresponding NO string.  The radius of the NO string is always taken
to be the distance between the center and the point where the field
strength is 25\% of the maximum.} At time $t=200$, the condition that
there be strings at least five times their width leads to the second
persistence line drawn in figure~\ref{persist}.

\pagebreak[3]

\EPSFIGURE[t]{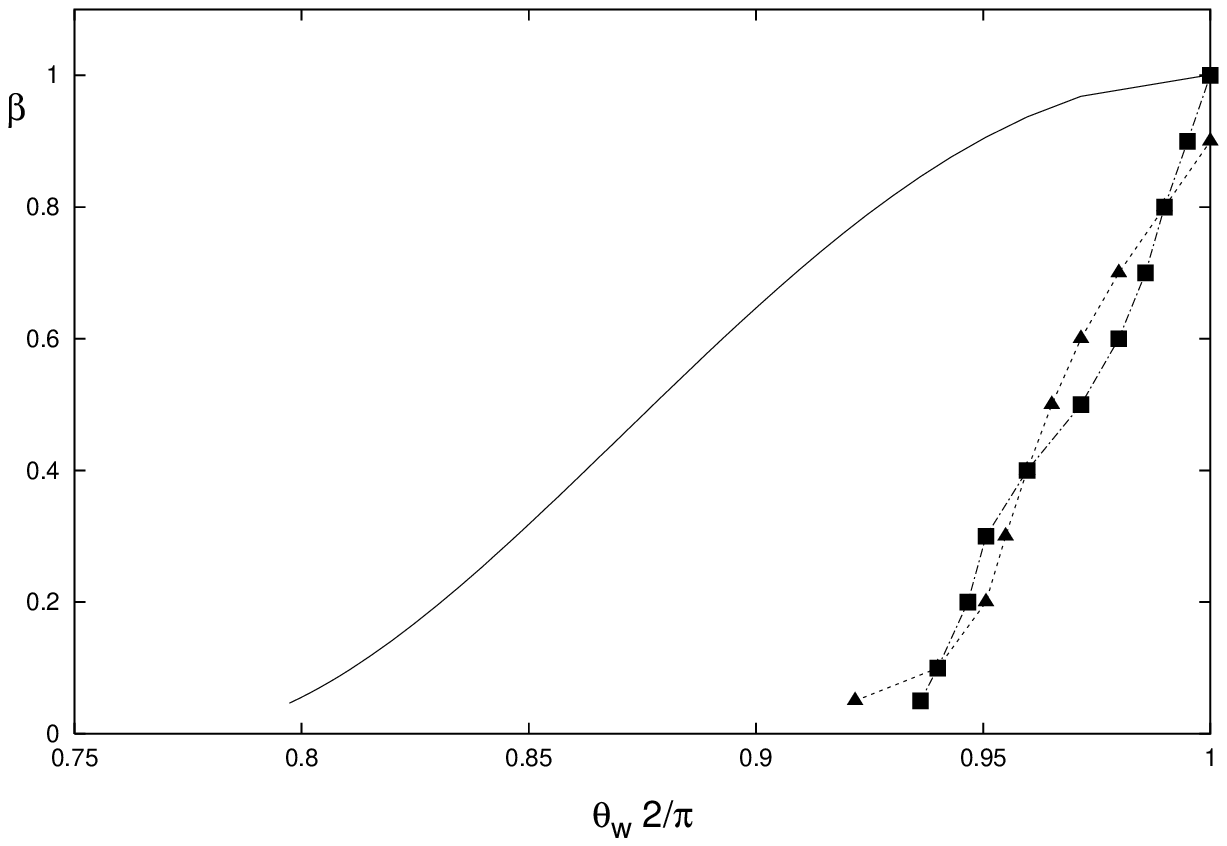,width=0.7\textwidth}{\label{persist}The solid line
is the semi-analytical curve marking the stability transition for
infinite, axially-symmetric $Z$-strings~\cite{JPV93}, and the points
mark the edge of the persistence region obtained in our simulations,
using both criteria described in the text. As expected, the
persistence limit lies entirely within the stability zone.  The
squares are obtained considering that the defects are persistent if at
$t=200$ there are more that 1000 points with a magnetic field strength
higher than 25\% of the maximum (see text); the triangles are obtained
by considering persistent those defects which at $t=200$ are at least
5 times longer than their width.}

\section{Conclusions}

The numerical simulations described here show that a significant
non-topological defect network can form in a generalized GSW model for
an electroweak phase transition.  The dynamics of such a network are
extremely complicated, driven by string segment linkings and by
isolated strings shrinking, and the details are highly sensitive to
the two model parameters $\tw$ and $\beta$.  Our principal result is
that in some regions of parameter space a persistent network of
genuine non-topological defects can form.  Though the actual version
of electroweak theory in our Universe lies outside this parameter
regime, these results show that, in models where topological (or
semilocal) defects are possible in some limit, it is possible to get a
network of non-topological defects close to that limit. Previous works
in the literature~\cite{V92,JPV93,BRT98} show that for parameters
close to those permitting topological (or semilocal) defects, there is
a regime where non-topological defects are stable. Our work confirms
this result, and shows that, although in a narrower region, a
sufficiently persistent network of defects can form in such a phase
transition.

Note that the generation of the string segments is intrinsically
dynamical and cannot be studied using initial condition arguments; in
particular our results are compatible with those of~\cite{NY96}.  Here
we are considering the time evolution of a network and not just the
initial configuration. In fact, the first timesteps in such
simulations correspond more to a numerical transient, in which
physically reasonable initial conditions are established, than an
actual phase transition.  Only after this initial transient can the
evolution of the network be trusted, and this is what determines
whether the defects persist or decay.  Given the primary role played
by the gauge fields, it would be interesting to know if our
conclusions generalize to non-topological string defects in other
models such as the two-Higgs standard model~\cite{BRT98,EJ93}.

\acknowledgments

\looseness=1 We thank Mark Hindmarsh and Tanmay Vachaspati for useful
discussions, and the referees for their comments.  AA and JU
acknowledge support from grants CICYT AEN99-0315 and UPV
063.310-EB187/98.  JU is also partially supported by a Marie Curie
Fellowship of the European Community programme HUMAN POTENTIAL under
contract number HPMT-CT-2000-00096. This research used resources of
the National Energy Research Scientific Computing Center, which is
supported by the Office of Science of the U.S. Department of Energy
under Contract No. DE-AC03-76SF00098. JU is grateful to the Lawrence
Berkeley National Laboratory at the University of California, the
Kapteyn Institute and the Institute for Theoretical Physics at the
University of Groningen, and the Astronomy Centre at the University of
Sussex for their hospitality during visits, and the use of their
computer facilities including those of the Sussex High-Performance
Computing Initiative.

{\renewcommand{\abovecaptionskip}{-6pt}
\FIGURE[t]{%
\epsfig{file=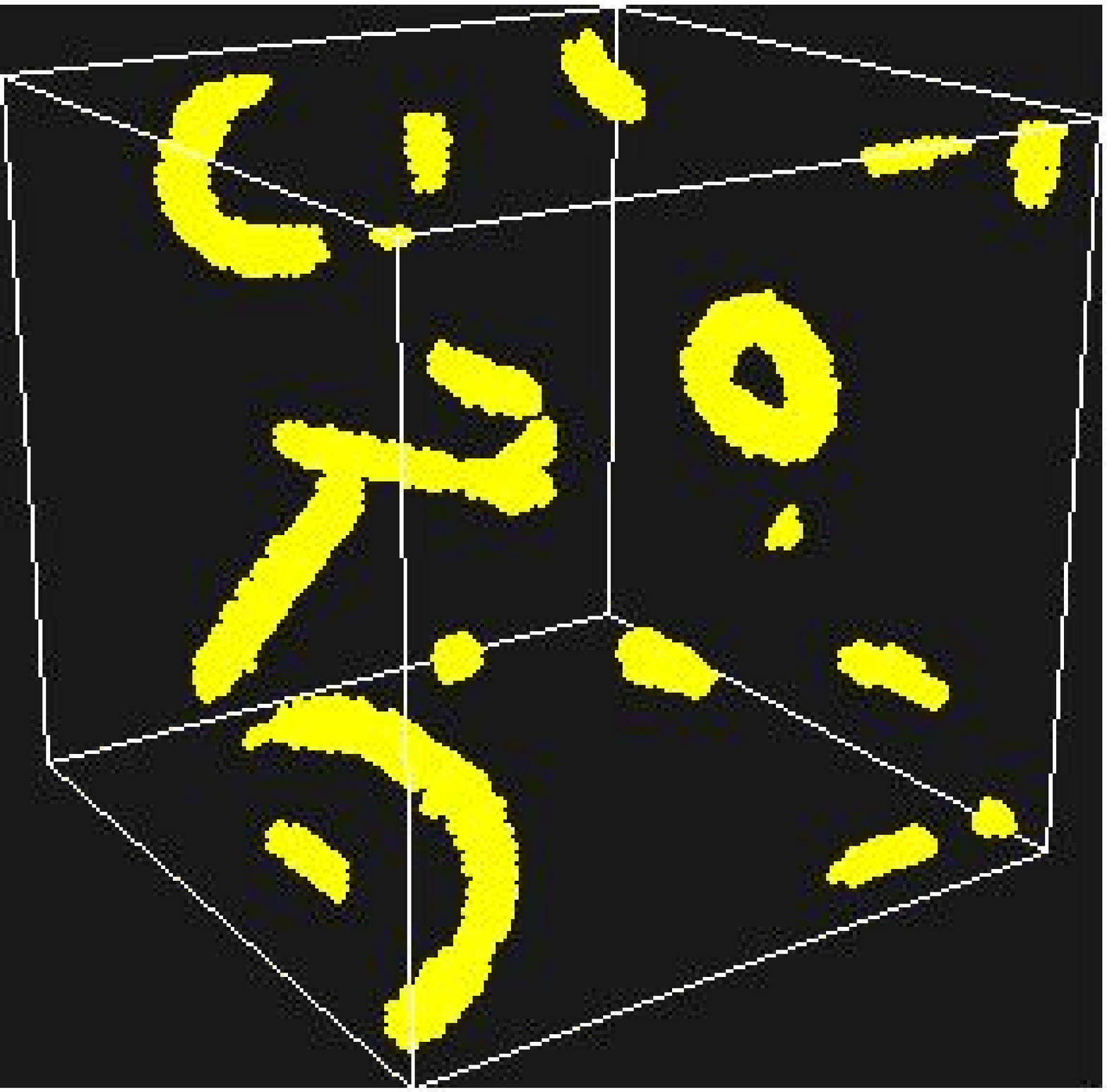, width=4.3cm} 
\epsfig{file=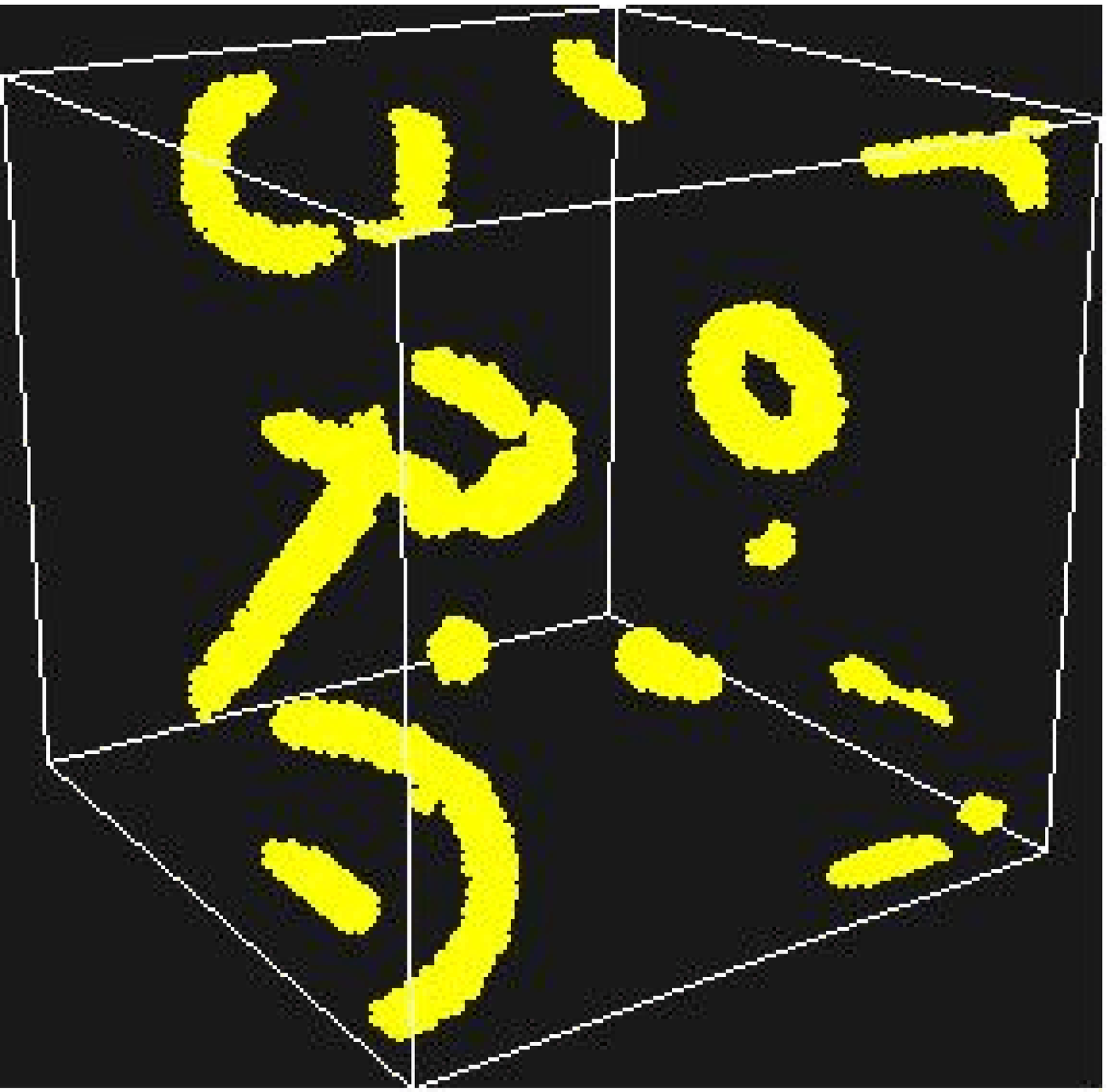, width=4.3cm}\linebreak
\epsfig{file=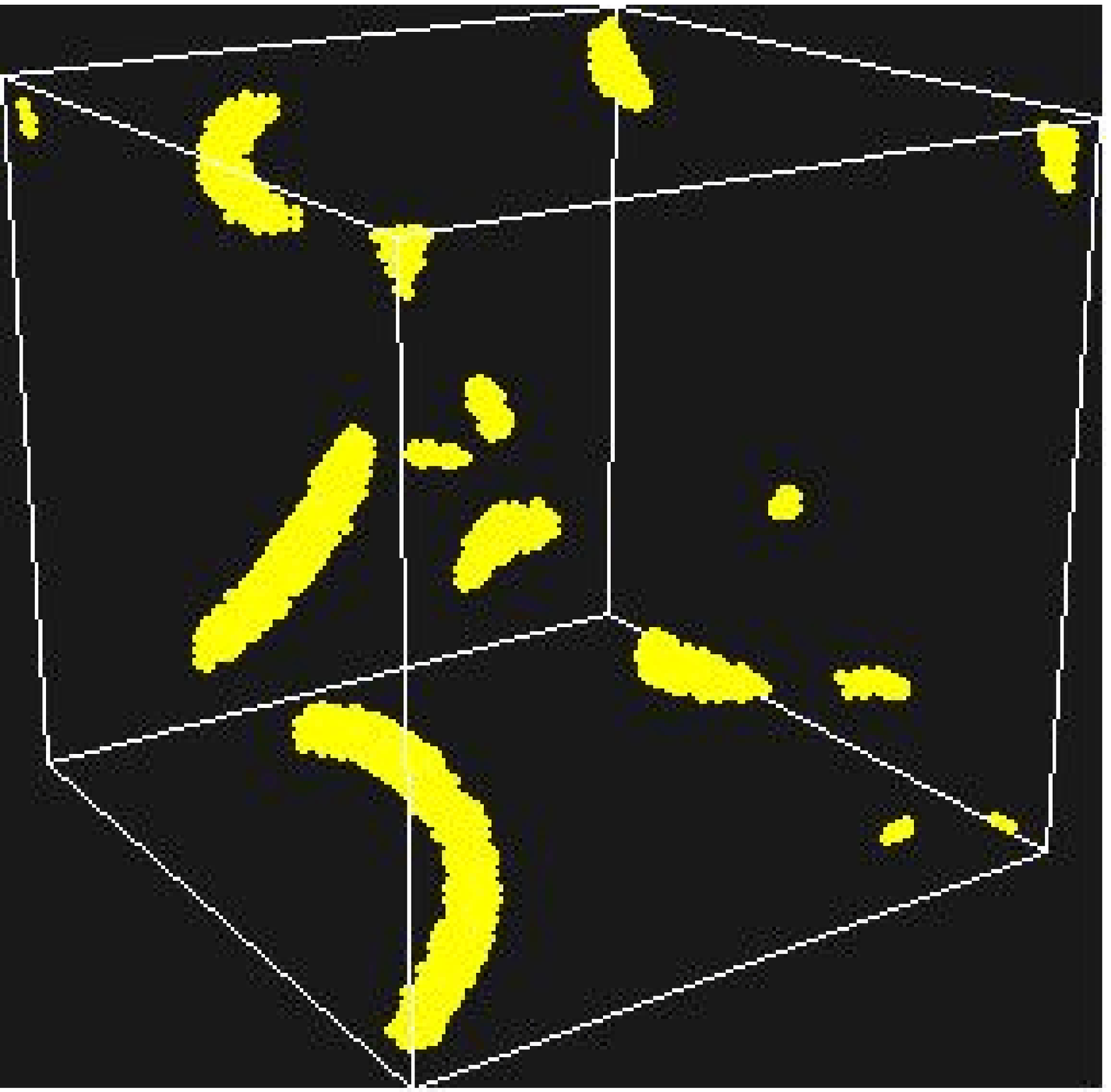, width=4.3cm} 
\epsfig{file=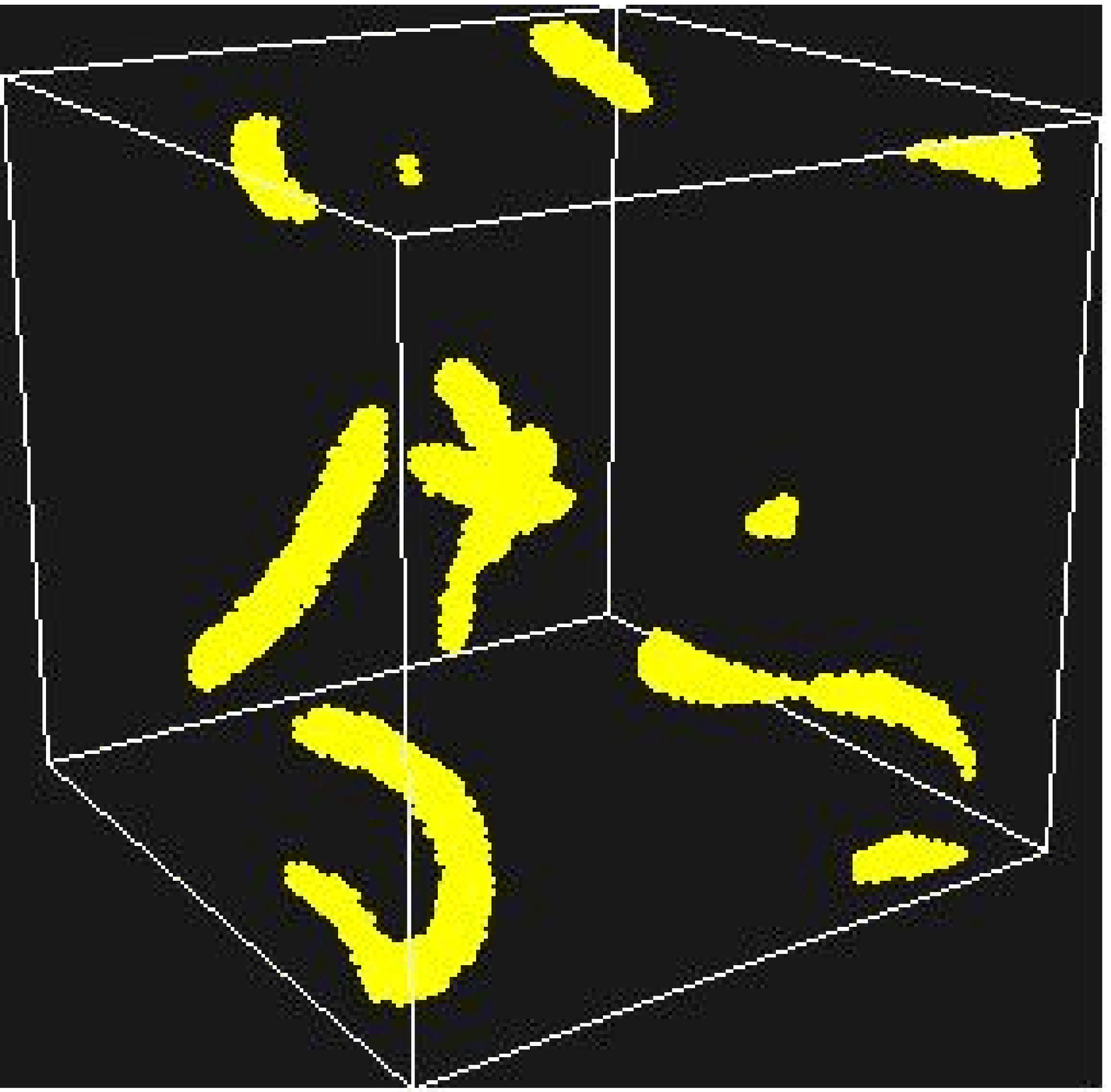, width=4.3cm}\linebreak
\caption[naive1]{\label{naive} Plot of the magnetic field strength for
the semilocal case at $t=40$ (top figures) and at $t=56$ (bottom
figures), using the same initial conditions, but evolving the system
using a na\"{\i}ve discretization (left) and a lattice link variable
discretization (right), as explained in the appendix. Both simulation
schemes are in very good agreement, with slightly longer strings in
the lattice variable discretization scheme.}}}

\appendix

\section{Discretization methods}

In this work, we discretized the equations of motion eq.~(\ref{eom})
replacing all the scalar and gauge fields by their values at the
lattice points. For instance, a $\U(1)$ covariant derivative was
substituted by
\be
D_j \phi(x)=\frac{\phi(x+x_j)-\phi(x-x_j)}{2 L}-
\frac{i g_{{\rm {\scriptscriptstyle Y}}}}{2}Y_i(x)\phi(x)
\ee
where $L$ is the lattice spacing.

It is well known that this discretization scheme does not respect
gauge invariance.  Nevertheless, we are dealing with well-defined
classical equations of motion in a particular gauge, and the breaking
of gauge invariance should be irrelevant in this context.  Indeed, we
make use of the fact that Gauss's Law is not automatically satisfied,
and by monitoring it during the evolution we can check that our
discretization is reasonably accurate.  This discretization was chosen
to facilitate comparison with earlier work~\cite{ABL98}.

There is an alternative discretization method widely used in the
literature which protects gauge invariance and recovers the original
equations of motion in the limit where the lattice spacing goes to
zero~\cite{KS}. This method uses lattice link variables, i.e., the
replacement of gauge fields by matrices living on the links between
lattice points. In this case, a $\U(1)$ covariant derivative will be
substituted by
\be
D_j\phi(x)=\frac{1}{L}\left(e^{-i L
Y_j(x)}\phi(x+x_j)-\phi(x)\right) .
\ee

In order to check whether our conclusions still hold using link
variables, we first performed a series of simulations in the semilocal
case to determine whether the evolution was compatible. Beginning from
the same initial conditions, we evolved the system according to each
discretization scheme. We observed that the simulations undergo
extremely similar evolution on a pointwise basis, as seen in
figure~\ref{naive}, though there are modest differences with slightly
longer strings on the lattice gauge calculations, occasionally leading
to extra connections.

\EPSFIGURE[t]{5.eps, angle=-90, width=0.6\textwidth}{\label{link}Plot
of string lengths in units of string widths (see text) for the
na\"{\i}ve discretization (squares) and the lattice link variable
formalism (circles). The data are measured at time $t=56$. The error
bars are 1 $\sigma$ over 10 simulations.}

To quantify this small difference and its impact on our statistical
result, we performed a further 120 simulations using both
discretizations. To save computing power we simulated our system in
the semilocal case ($\sin^2\tw=1$), discretized in $64^3$ cubes for
different values of $\beta$.  We used several different initial
conditions, but for each one we let the system evolve using both the
na\"{\i}ve discretization of the (lagrangian) equations of motion and
the link variable method for discretizing the (hamiltonian) equations
of motion.  The initial conditions were calculated using the method b)
described in the text (section~\ref{ns}), and as in the electroweak
simulations performed in this work, we added an \emph{ad hoc} damping
term ($\eta=0.5$). As our interest lies mainly in the late-time
behaviour of the dumbbell network, because the smaller cube size
forces $t\lesssim 64$, we chose to calculate the total string length
at $t=56$ for both cases.
  
Figure~\ref{link} shows the results.  We computed the string lengths
as explained in section~\ref{results}, footnote~\ref{automate}, and
then added all string lengths for strings longer than five times their
length.  That final measure is represented in figure~\ref{link}, which
shows clearly that the differences between the schemes are well within
the uncertainties.  Although performed for only $64^3$ cubes, this
result should continue to hold for the larger cubes used to derive the
main results in the body of this paper. The system simulated in this
test is only the semilocal case, not the full electroweak case, but we
are convinced that using the lattice link variable method will not
alter significantly the results presented in this paper.

\end{document}